\documentclass[conference, a4paper]{IEEEtran}
\usepackage{geometry}
\usepackage[utf8]{inputenc}
\usepackage{amsmath,amssymb,amsfonts}
\usepackage{textgreek}
\usepackage{graphicx}
\usepackage{color,bm}
\usepackage{wasysym}
\usepackage{tikz}
\usepackage{cite}
 \usepackage{array,multirow}
 \usepackage{float}
 \usepackage{subfig}
 \usepackage{balance}
 
\ifCLASSINFOpdf
\else
\fi
\usepackage{graphicx}
\usepackage{todonotes}
\DeclareGraphicsExtensions{.tif}
\usepackage{dblfloatfix}
\usepackage{url}
\begin{document}
%
\title{\LARGE \textbf{Early Findings from Field Trials of Heavy-Duty Truck Connected Eco-Driving System}}

\author{\IEEEauthorblockN{
Ziran Wang\IEEEauthorrefmark{1},
Yuan-Pu Hsu\IEEEauthorrefmark{1},
Alexander Vu\IEEEauthorrefmark{1},
Francisco Caballero\IEEEauthorrefmark{1},\\
Peng Hao\IEEEauthorrefmark{1}, 
Guoyuan Wu\IEEEauthorrefmark{1},
Kanok Boriboonsomsin\IEEEauthorrefmark{1},
Matthew J. Barth\IEEEauthorrefmark{1},\\
Aravind Kailas\IEEEauthorrefmark{2},
Pascal Amar\IEEEauthorrefmark{2},
Eddie Garmon\IEEEauthorrefmark{2} and
Sandeep Tanugula\IEEEauthorrefmark{2}}

\IEEEauthorblockA{\IEEEauthorrefmark{1}College of Engineering - Center for Environmental
Research and Technology\\
University of California, Riverside, CA, USA 92521}
\IEEEauthorblockA{\IEEEauthorrefmark{2}Volvo Group North America\\ Greensboro, NC, USA 27409\\
Email: kanok@cert.ucr.edu}}

\maketitle

\begin{abstract}
In recent years, the development of connected and automated vehicle (CAV) technology has inspired numerous advanced applications targeted at improving existing transportation systems. As one of the widely studied applications of CAV technology, connected eco-driving takes advantage of Signal Phase and Timing (SPaT) information from traffic signals to enable CAVs to approach and depart from signalized intersections in an energy-efficient manner. However, the majority of the connected eco-driving studies have been numerical or microscopic traffic simulations. Only few studies have implemented the application on real vehicles, and even fewer have been focused on heavy-duty trucks. In this study, we developed a connected eco-driving system and equipped it on a heavy-duty diesel truck using cellular-based wireless communications. Field trials were conducted in the City of Carson, California, along two corridors with six connected signalized intersections capable of communicating their SPaT information. Early results showed the benefits of the system in smoothing the speed profiles of the equipped truck when approaching the connected signalized intersections.\end{abstract}

\IEEEpeerreviewmaketitle

\section{Introduction}
\IEEEPARstart
In recent years, transportation activities continue to have significant impacts on energy consumption and air quality. In 2017, about 29\% of U.S. energy consumption was for transporting people and goods from one place to another \cite{USEIA2017}. Over 90\% of the fuel used for transportation is petroleum based, including primarily gasoline and diesel. Transportation-related greenhouse gas (GHG) emissions was the second largest contributor to the nationwide GHG emissions inventory in 2017, accounting for approximately 28.9\% of total U.S. GHG emissions \cite{USEPA2017}. In 2018, the transportation sector exceeded the power sector as the largest contributor of $CO_2$ emissions in the U.S. \cite{Rhodium2019}.

These facts have been increasing public awareness of the need to reduce energy consumption and pollutant emissions generated by our contemporary transportation system. Among all the strategies to mitigate the transportation-related environmental footprint, eco-driving along urban arterial has gained significant research interest around the world \cite{barth2009energy, kamal2010on, saboohi2009model, mensing2011vehicle, wang2018cluster}. By applying connected and automated vehicle (CAV) technology, drivers would effectively reduce the number of full stops and idling time, and avoid unnecessary acceleration and deceleration by receiving downstream traffic information, such as queue length and signal phase and timing (SPaT). The Eco-Approach and Departure (EAD) application is a typical example, where drivers are guided to approach and depart from signalized intersections in an environmentally friendly manner using SPaT and geometric intersection description (GID) information sent from the wireless roadside units (RSU) installed as part of the signalized intersection infrastructure \cite{xia2013development}, \cite{barth2011dynamic}. In such a manner, CAVs can increase their energy efficiency and decrease pollutant emissions by simply following well-designed speed profiles while traveling through the intersection. Results of microscopic simulation models showed a 10-15\% reduction on energy consumption and CO2 emissions by applying the EAD application to fixed-timing signalized intersections \cite{xia2013dynamic}. In congested traffic conditions, the EAD application also worked effectively where preceding queues could be estimated by leveraging real time vehicle detection and signal information system \cite{he2015optimal}. Moreover, it was also revealed in previous studies that, drivers’ behavior adaptability under actual driving conditions also played an important role in the effectiveness of the EAD application \cite{xiang2015aclosed}.

Many studies have been focused on developing eco-driving systems and evaluating the proposed systems by numerical or microscopic simulations. However, only a few implemented the proposed technology on real vehicles. A field test along the El Camino Real corridor in Palo Alto, CA was conducted to evaluate the EAD application for \textit{actuated signals} in real-world traffic\cite{hao2019eco}. The tested track is 1.7 miles long, and all intersections are equipped with DSRC RSUs. A test vehicle (2.5 L 4-cylinder) was equipped with an industry-grade radar, GPS, DSRC modem, on-board computer, and an artificial dashboard. It was shown up to 6\% energy and emissions reduction can be achieved in real-world traffic. Additionally, a partially automated (longitudinal control) EAD prototype was developed and demonstrated at the Turner-Fairbank Highway Research Center (TFHRC) in McLean, Virginia, and the results showed an average of 17\% in energy consumption reduction \cite{altan2017glidepath}.   

\newpage

\begin{figure*}[!b]
    \centering
    \includegraphics[width=2.0\columnwidth]{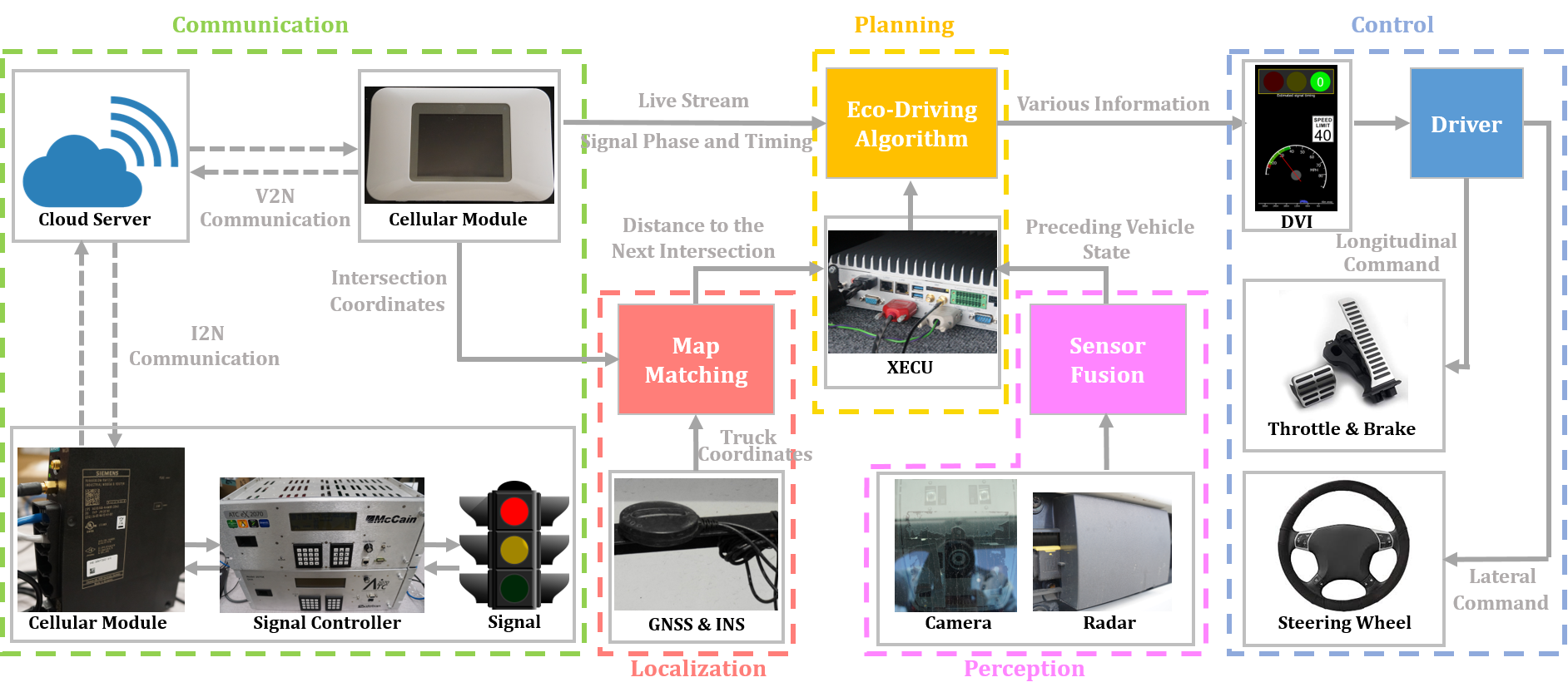}
    \caption{Architecture of the connected eco-driving system for heavy-duty trucks.}
    \label{flowchart}
\end{figure*} 

Built on top of the authors' previous work on light-duty vehicles, a connected eco-driving system for heavy-duty trucks has been developed in this study. Since heavy-duty trucks play a significant role in transportation-related (especially on goods movement) energy consumption and pollutant emissions, we have adapted the existing eco-driving system to heavy-duty trucks to better shape our contemporary transportation system. Furthermore, different from our previous implementation with DSRC, we have used 4G-LTE to enable cellular-based wireless communications in this study, given its benefits of scalability and wider coverage. We have implemented a simplified eco-driving algorithm to calculate the recommended speed for the test truck in real time. The algorithm has been strategically integrated with other system components, such as Global Navigation Satellite Systems (GNSS), on-bard maps, radar, X Electronic Control Unit (XECU), and live SPaT streams. A user-friendly driver-vehicle interface (DVI) has been designed, so the driver can follow the recommended speed to drive the truck to achieve the goal of energy savings.  A field test of the developed connected eco-driving system has been conducted in Carson, California with a heavy-duty diesel truck, along a test route with six equipped signalized intersections capable of broadcasting their SPaT information.  

The remainder of this paper is organized as follows: Section II demonstrates the architecture of the proposed connected eco-driving system for heavy-duty trucks. Section III elaborates two key modules: map matching and eco-driving algorithm of the proposed system. The field testing conducted in Carson, CA is presented in Section IV, followed by result analysis. Section V concludes this paper along with the discussions on future work.

\section{Architecture of the Connected Eco-Driving System}
In this section, the architecture of the proposed connected eco-driving system is introduced, with illustrations of both software and hardware components installed. As can be seen from Fig. \ref{flowchart}, the whole system is divided into five different modules: communication, localization, perception, planning, and control, whose respective functions are elaborated below.

\subsection{Communication Module}
The communication module facilitates real-time and reliable \textit{vehicle-to-network} (V2N) and \textit{ infrastructure-to-network} (I2N) communications. Although short-range communication technology such as DSRC is widely used by various CAV applications, their low scalability and asynchronous disadvantage may lead to several issues. As a substitute, 4G-LTE has been adopted to enhance V2N and I2N communications given its versatile communications types from one-to-one to one-to-multiple transmissions, with both simulations and field experiments conducted by researchers all over the world during the past decade \cite{xu2017DSRC, nshimiyimana2016comprehensive}. In our connected eco-driving system, we also install a 4G-LTE cellular module with AT\&T SIM card on board to enable V2N communication. Note that, compared to DSRC, one major disadvantage of 4G-LTE is its relatively longer latency. However, based on the our findings in our field implementation of this study, the latency introduced by 4G-LTE is about 100 ms, which is satisfactory for the connected eco-driving system.

Each signal head on the test track is controlled by a McCain 2070E controller, which is a rugged, multi-tasking field process and communication system configurable for a variety of traffic management applications. Additionally, each controller is equipped with a SIEMENS RM1224 mobile wireless router for 4G-LTE connection, so the raw SPaT data could be sent to the cloud server through I2N communication. Then the 4G-LTE cellular module on the truck can request the live stream of SPaT information from the cloud server at every second.

\subsection{Localization Module}
The localization module of the connected eco-driving system consists of hardware component, a global navigation satellite system (GNSS) and inertial navigation system (INS), and software component, a map matching algorithm. The GNSS \& INS component serves as a combined satellite \& inertial-based navigation system, which can be optionally augmented by terrestrial reference stations \cite{geiger2012team}. This component can provide precise position, movement, and posture measurements for the self-localization and altitude determination of the truck by differential correction. On the test truck, a versatile U-BLOX NEO-M8N GNSS module is installed to locate the truck's position. This GNSS module has a navigation sensitivity of -167dBm, which provides outstanding positioning accuracy in scenarios where urban canyon or weak signals are involved.

For the map matching component, a pre-built map of the testing field is available on board, with information such as the intersection ID, signal group ID, signal location, road type, road name and direction, and road speed limit. The truck's coordinates (\textit{i.e.}, longitude, latitude, and heading) received from the GNSS \& INS component can then be matched to the pre-built map by the proposed map matching algorithm, where the details of the algorithm will be introduced in Sec. \ref{sec:algorithms}A.

\subsection{Perception Module}
Different from the communication module that provides information about the signal status (\textit{i.e.}, SPaT information), the perception module can provide information about other surrounding vehicles (especially if they are not CAVs). In this connected eco-driving system , we adopt the forward sensing kit to detect any preceding vehicles, bicycles, and/or pedestrians. The Volvo EU Production Radar Unit together with the forward-looking camera can detect the movement of preceding vehicle, and send its dynamics information to the sensor fusion component. Once the detected objects are classified by the sensor fusion component, the desired information is sent to the eco-driving algorithm. Given the measured relative speed and distance between the truck and its preceding vehicle, the time-to-collision value can be computed. If the time-to-collision value is lower than a preset threshold, the suggested speed for the truck driver will be turned off on the DVI, and the driver will drive the truck without any auxiliary information. 

\subsection{Planning Module}
The planning module is the core of the proposed connected eco-driving system. The XECU of the truck is equipped with a ruggedized computer, which has access to multiple channels of CAN data from different sub-networks of the truck. It receives the distance to the next signal from the localization module, preceding vehicles' states from the perception module, and the speed of the truck from CAN message.

The eco-driving algorithm is embedded in the XECU, based on a Microsoft .NET 4.7 framework with C\# language. Data query, recommended speed calculation, and the signal display are all updated at 10 Hz frequency to achieve the consistency, while the communication module is operating at 1 Hz due to the restriction on infrastructure. All the concurrent tasks are handled by threading in the main thread with shared memory. Upon receiving all this information, the recommended speed of the truck at the next time step for the driver will be computed, and sent to the DVI along with some other information. On the other hand, we also utilize a package called log4net for archiving the raw data and the computation results. The details of the eco-driving algorithm will be introduced in Sec. \ref{sec:algorithms}B.

\subsection{Control Module}
Once the recommended speed is computed by the eco-driving algorithm in the planning module, the control module of the connected eco-driving system starts to play. A DVI is shown on the monitor right above the dashboard of the truck. With the auxiliary help from the DVI, the driver controls the longitudinal movement of the truck via the throttle and brake, and controls the lateral movement  by steering wheel. 

\begin{figure}[h]
    \centering
    \includegraphics[width=0.32\columnwidth]{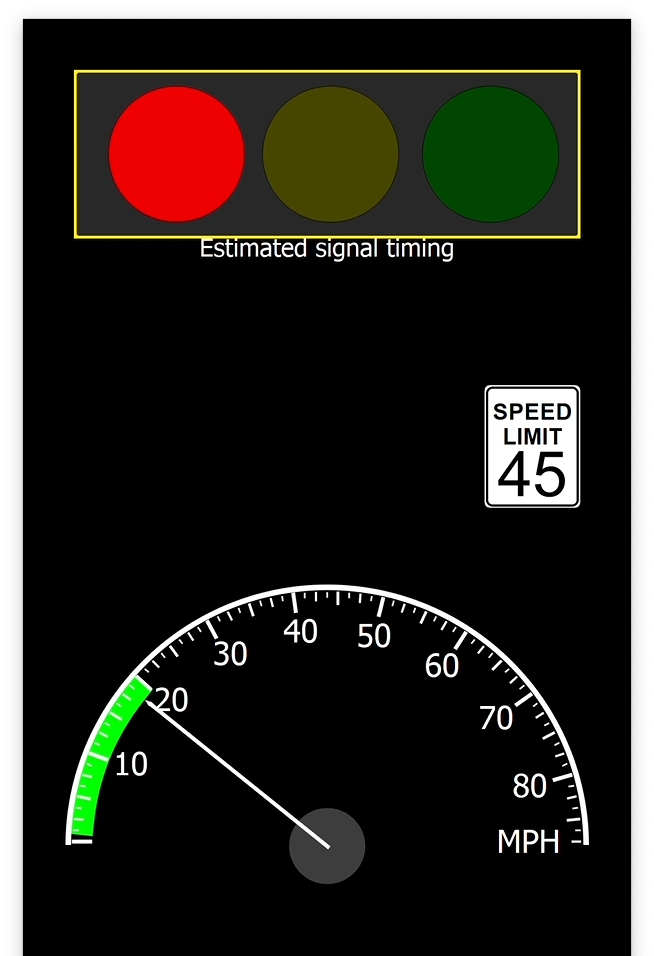}
    \includegraphics[width=0.32\columnwidth]{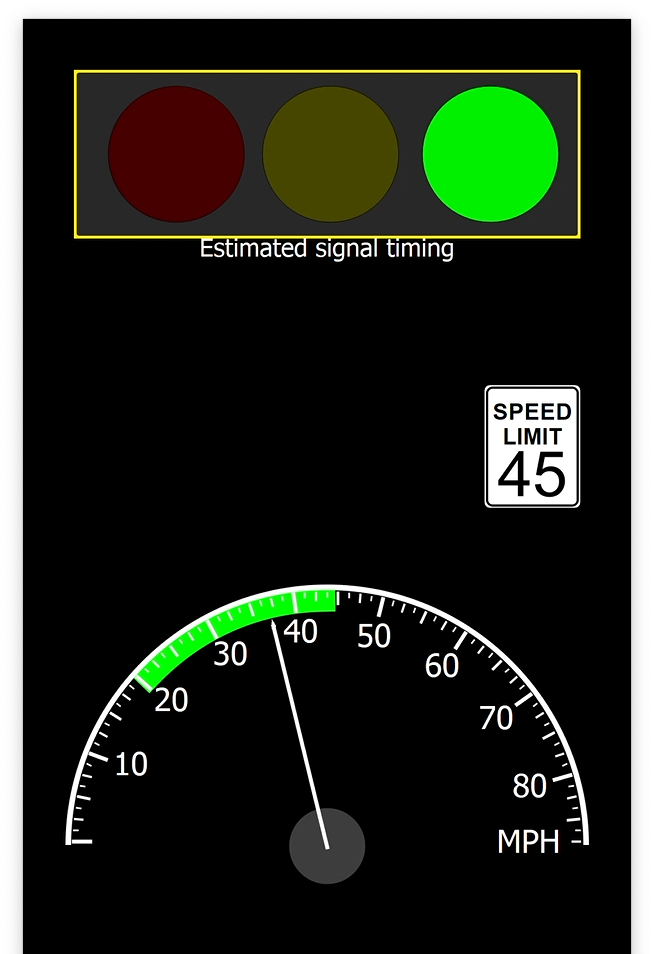}
    \includegraphics[width=0.32\columnwidth]{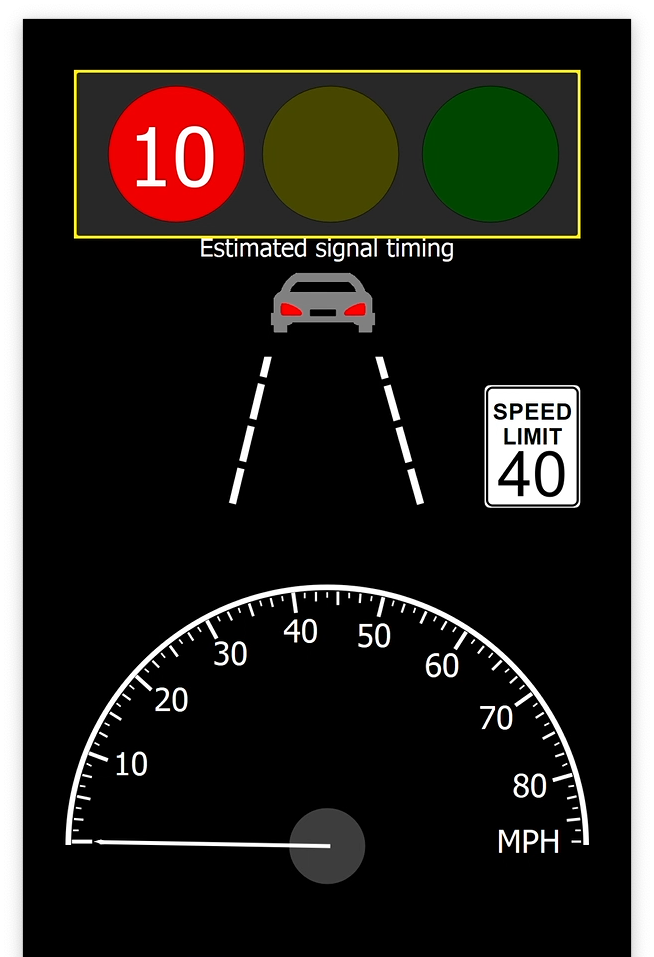}
    \caption{Screen shots of DVI designed for the truck driver.}
    \label{DVI}
\end{figure}

Three typical scenarios are shown in Fig. \ref{DVI}. As can be seen from these subplots, the top of the DVI shows the upcoming signal phase, but not necessarily the signal timing. We only show the signal timing when the truck is at full stop (shown as the right subplot), but hide that information in other scenarios (the left and middle subplot) to avoid distraction to the driver while driving the truck. The right hand side of the DVI is the speed limit of the current roadway segment, which is acquired from the pre-built map. The artificial speedometer is put at the bottom of the DVI with a green band to demonstrate the recommended speed range for the truck driver. As can be seen from the right subplot, if a vehicle is detected in front of the truck within a certain range, a vehicle icon is shown in the middle of the DVI, and the upper bound of the recommended speed is bounded by the speed of the preceding vehicle. If the computed time-to-collision value is lower than a preset threshold, the DVI would display a warning message instead of showing the SPaT information, allowing the driver to disengage the eco-driving system and be aware of potential rear-end collision by himself/herself.

\section{Algorithms of the Connected Eco-Driving System} \label{sec:algorithms}
In this section, we briefly introduce two crucial algorithms in the developed connected eco-driving system: The map matching algorithm in the localization module, and the eco-driving algorithm in the planning module.

\subsection{Map Matching Algorithm}
The first step of our map matching algorithm is to find the projection point of truck's current latitude and longitude on the nearest lane of the hard-coded map. Since we calculate the distance on the surface of the earth, Haversine formula can be applied as:
\begin{equation}\small
    d = 2r \cdot sin^{-1}(\sqrt{sin^2(\frac{\Delta\varphi}{2}) + cos(\varphi_1)cos(\varphi_2)sin^2(\frac{\Delta\lambda}{2})})
\end{equation}
where $d$ is the distance between two points on the sphere, $r$ is the radius of the earth, $\varphi_1$, $\varphi_2$, $\Delta\varphi$ represent the latitude of the first point, the latitude of the second point, and  the difference between $\varphi_1$ and $\varphi_2$, respectively, $\Delta\lambda$ is the longitude difference between the two points.

Since the direction to the next signal should be consistent with the current heading of the truck, we verify this procedure by using inner product:

\begin{equation}
\hat{h} \cdot \hat{u} > 0\\
\end{equation}
where $\hat{h}$ is the current heading of the truck and $\hat{u}$ is the direction of the intersection. 

Given that the GNSS signal might drift from the current lane of the road, errors could be generated and accumulated if we directly compute the distance to the intersection node. Therefore, we utilize the triangle formed by the truck's coordinate and its two adjacent nodes' coordinates to calculate the real distance from the projection point of the truck to the intersection. Heron's formula can be given as:

\begin{equation}
A = \sqrt{s(s-d_{tn1})(s-d_{tn2})(s-d_{n1n2})}
\end{equation} 
where $A$ is the area of the triangle whose sides have lengths $d_{tn1}$ (truck to the previous node), $d_{tn2}$ (truck to the next node), and $d_{n1n2}$ (the previous node to the next node), and $s$ is the semi-perimeter of the triangle defined as

\begin{equation}
s = \frac{d_{tn1} + d_{tn2} + d_{n1n2}}{2}
\end{equation}

Then, we can compute the height $h$ of the triangle by the area and the base-side $d_{n1n2}$, where $h$ is considered as the error distance due to the drift of GNSS signal from the current lane. Based on $h$ and $d_{tn2}$, we could use Pythagorean theorem to compute the distance $d_{pn2}$ from the projection point of the truck to the downstream node. Finally, the distance from the truck to the next signal $d_{sig}$ can be calculated by adding the resulted distance with the lengths of all downstream segments before the signal:
\begin{equation}
d_{sig} = d_{pn2} + d_{n2n3} + d_{n3n4} + ... + d_{n(m-1)n(m)}
\end{equation}
where $n(m)$ is the node with respect to the next signal.

\subsection{Eco-Driving Algorithm}
Although we developed several versions of eco-driving algorithms before (\textit{e.g.}, piece-wise trigonometric algorithm \cite{xia2013development, xia2013dynamic}, graph-based algorithm\cite{Jin2016ecoacc} \cite{hao2018connected}), in this system, we develop a simplified version of eco-driving algorithm for two major reasons:
\begin{itemize}
    \item Compared to previous eco-driving algorithms, this simplified version of algorithm is easier to implement on real vehicles, and also more attractive in terms of real-time performance;
    \item This version of algorithm computes a feasible speed range instead of a single value, and provides an easy-to-follow speed band on the DVI for the truck driver. Since the ability to precisely follow the recommended speed is limited and varied by the driver, and the response time of the driver to maneuver the truck may be longer than light duty vehicles, the feasible speed range option should be more robust and user-friendly for truck drivers.
\end{itemize} 
In this subsection, we demonstrate the algorithm with parameters $d_{sig}$ as distance to the intersection, $t_{current}$ as the time left of current signal, and $v_{lim}$ as the speed limit of the roadway. The reference speed $v_0$ of the truck can then be given as
\begin{equation}
v_0 = d_{sig}/t_{current}
\end{equation}

We then categorize the eco-driving algorithm into two different scenarios based on the current signal phase:

\subsubsection{Red Phase}
\begin{equation}
    \left\{
                \begin{array}{ll}
                  v_{lower} = 0, v_{upper} = v_0 & \textit{, if } v_0 \leq v_{lim}\\
                  v_{lower} = 0, v_{upper} = v_{lim} & \textit{, otherwise}\\
                \end{array}
              \right.
\end{equation}

\noindent When the signal phase at the upcoming intersection is red, the lower bound of the suggested speed $v_{lower}$ would always be zero. However, the upper bound $v_{upper}$ would be set to the speed that the truck can pass the intersection just at the moment it turns green without stopping. The highest speed is also constrained by the speed limit $v_{lim}$ of the roadway.

\subsubsection{Green Phase}
\begin{equation}
    \left\{
                \begin{array}{ll}
                  v_{lower} = v_0, v_{upper} = v_{lim} & \textit{, if } v_0 \leq v_{lim}\\
                  v_{lower} = 0, v_{upper} = 0 & \textit{, otherwise}\\
                \end{array}
              \right.
\end{equation}

\noindent On the other hand, when the signal at the upcoming intersection is green, the upper bound $v_{upper}$ of the recommendation would be the speed limit, and the lower bound $v_{lower}$ is set to the lowest speed that the truck can pass the intersection during the current green phase. Under the circumstances that even using the highest speed (\textit{i.e.}, speed limit) the truck still cannot pass the intersection, the algorithm will suggest the driver to slow down and wait for the next green phase.

\section{Field Implementation}

\begin{figure}[ht]
    \centering
    \includegraphics[width=1.0\columnwidth]{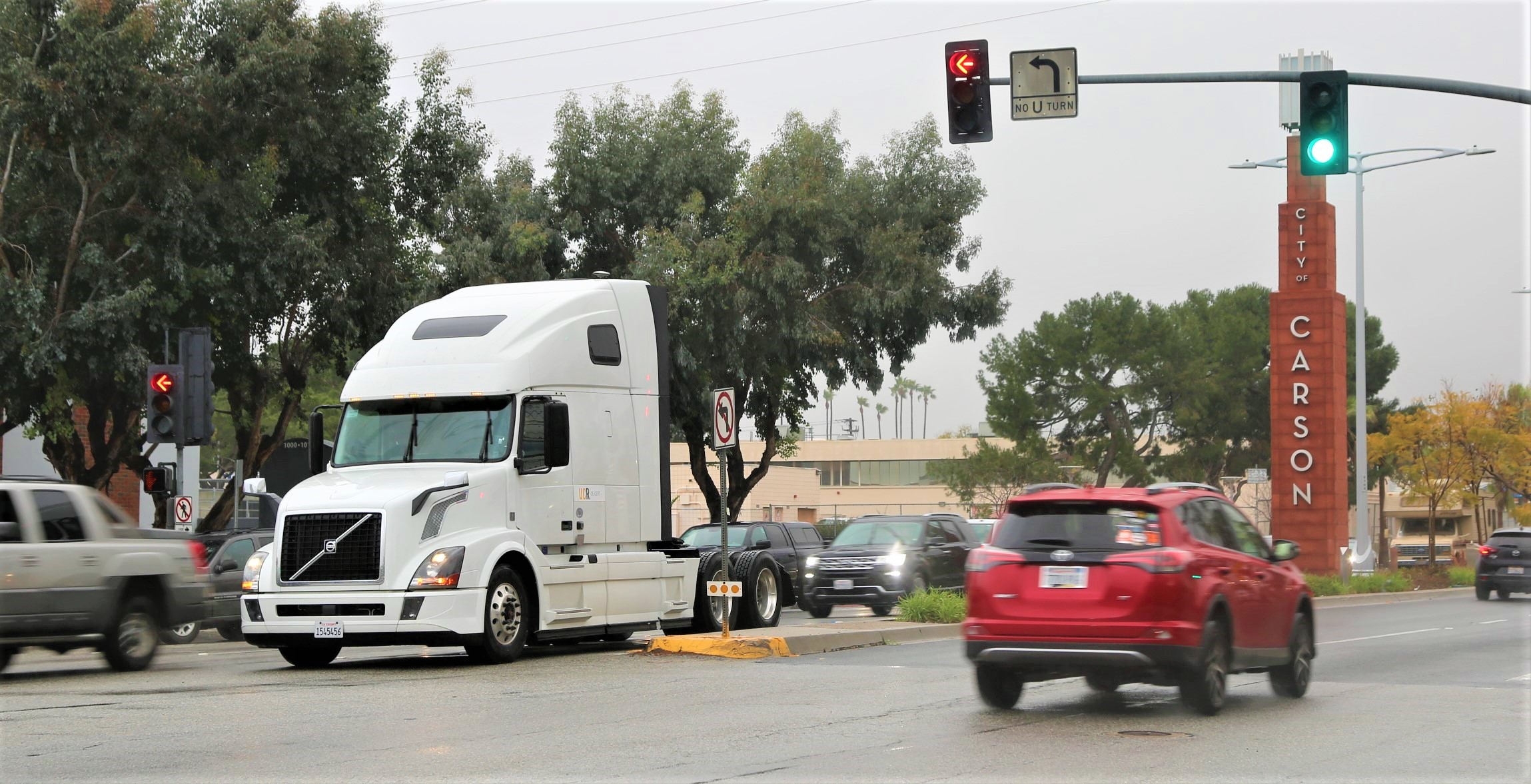}
    \caption{Field implementation is conducted by the heavy-duty truck in Carson, California.}
    \label{Truck}
\end{figure}
\begin{figure}[ht]
    \centering
    \includegraphics[width=1.0\columnwidth]{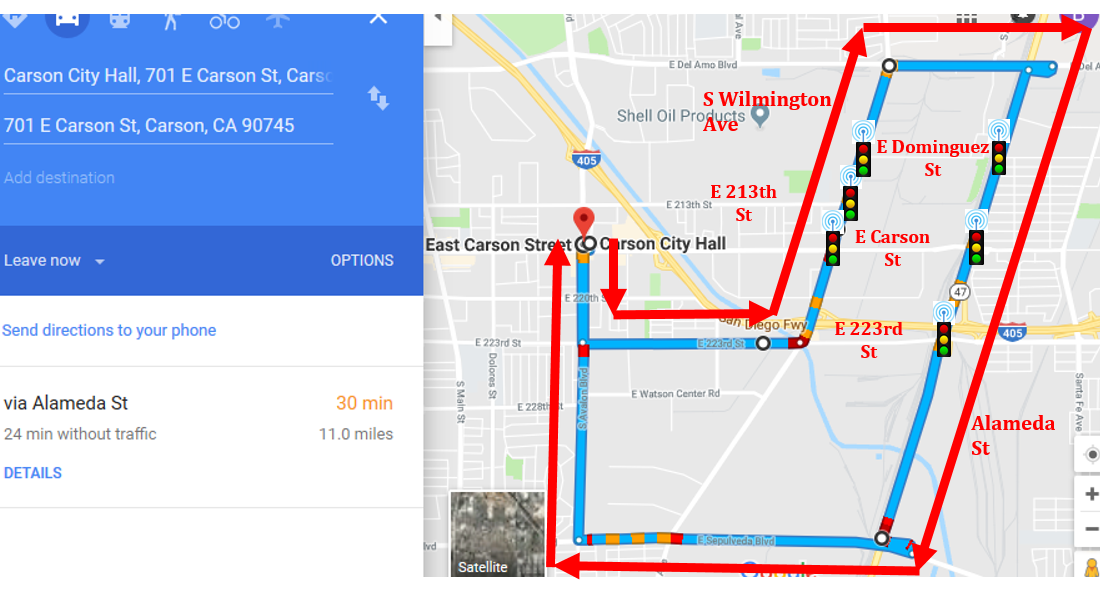}
    \caption{Field implementation route in Carson, CA with six equipped signalized intersections.}
    \label{Route}
\end{figure}

In this work, we adopt a Volvo VNL heavy-duty diesel truck for our field implementation, which is a 2015 Volvo VNL model with a 13L diesel engine (455 HP) and a maximum cruise speed of 115 km/h and is shown in Fig. \ref{Truck}. The test runs are conducted in Carson, CA, which locates right next to the Port of Los Angeles and Port of Long Beach, two busiest container ports in the United States. Numerous factories and warehouses are located in the testing area, so the proposed connected eco-driving system can potentially save energy consumed by heavy-duty trucks travelled in this area. Fig. \ref{Route} shows the test route where six signalized intersections on Wilmington Blvd. and Alameda St. are equipped with communication modules. The respective SPaT information can be sent to the cloud and allow our connected truck to fetch the data.

\begin{figure}[ht]
    \centering
    \includegraphics[width=1.0\columnwidth]{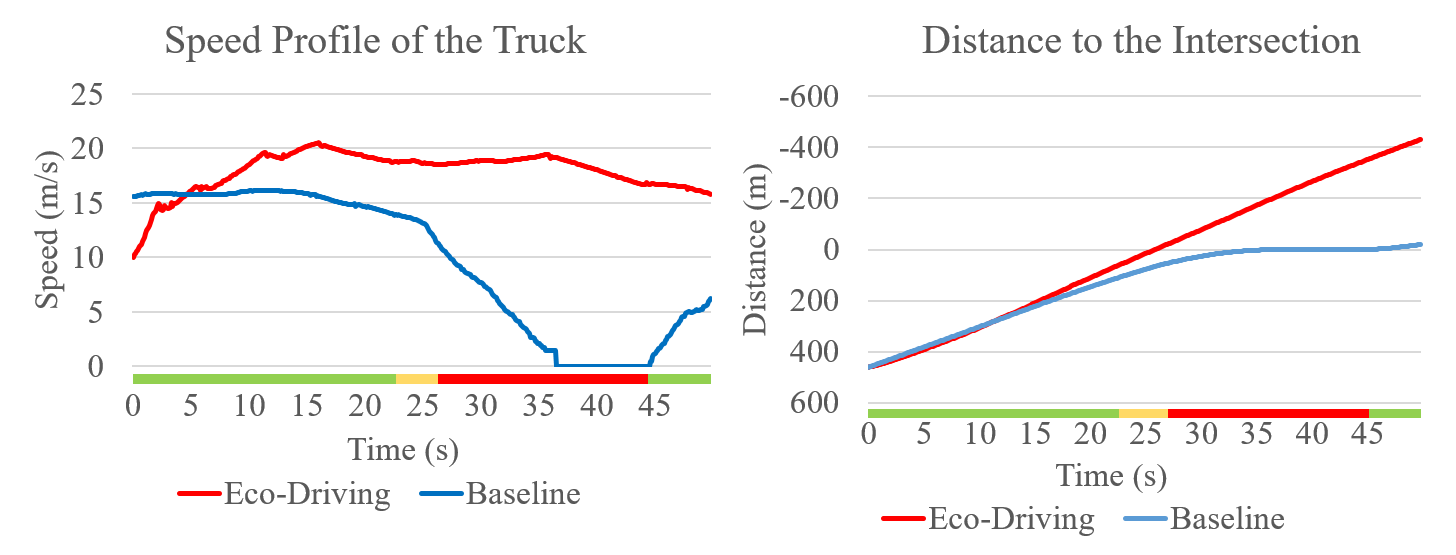}
    \caption{Test results of the acceleration scenario of the connected eco-driving vs. the baseline.}
    \label{Acceleration}
\end{figure}

\begin{figure}[ht]
    \centering
    \includegraphics[width=1.0\columnwidth]{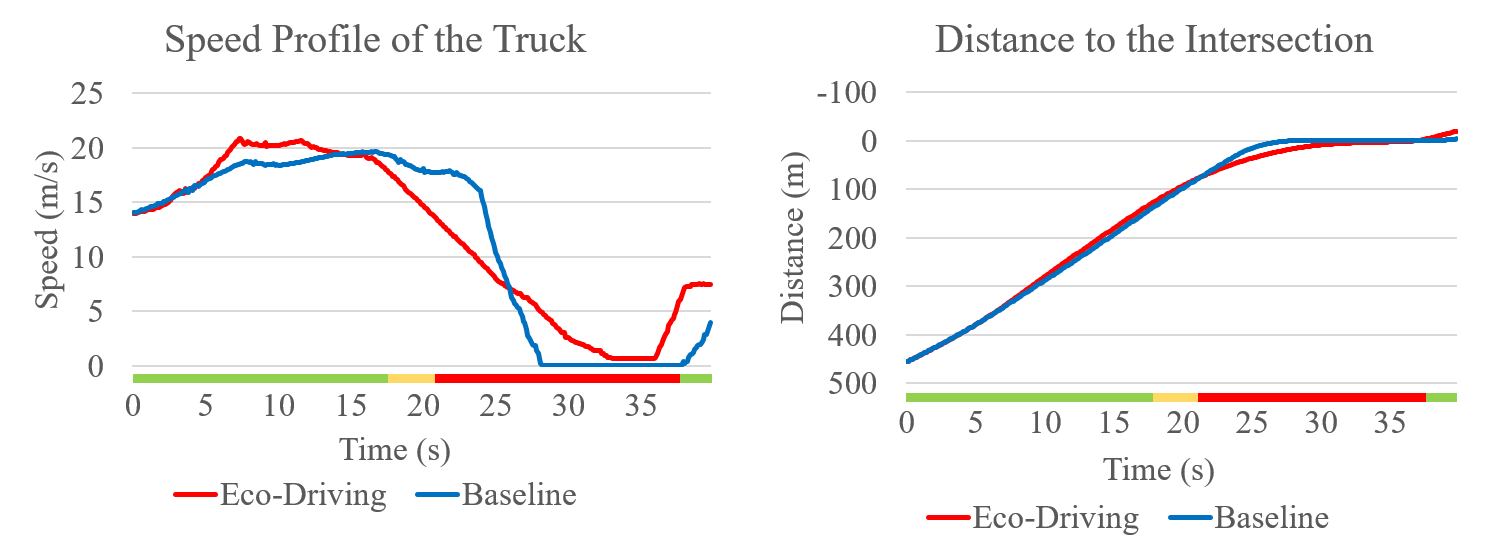}
    \caption{Test results of the deceleration scenario of the connected eco-driving vs. the baseline.}
    \label{Deceleration}
\end{figure}

For a comparative study, the truck ran a few times along the test route with and without the developed connected eco-driving system. As it was driven in real traffic environment, where the traffic flow may vary over time of the day, it is not fair to quantitatively compare the energy consumption between different test runs. Therefore, in this section of the paper, we select results generated during four test runs, where two of them are with connected eco-driving system and the other two are without it (\textit{i.e.,} baseline). To make this comparison as fair as possible, all four selected test runs are conducted at the same intersection with clear downstream traffic, which means the movement of the truck is not affected by other vehicles on the road. Also, in each scenario shown in Fig. \ref{Acceleration} and Fig. \ref{Deceleration}, the truck approaches the intersection from the same distance with the same SPaT status (shown as the color bar below the x axis). Therefore, the difference between two trajectories (eco-driving and baseline) in each plot is mainly caused by the existence of the proposed connected eco-driving system.

Fig. \ref{Acceleration} shows the comparison results on the acceleration scenario between the connected eco-driving system and the baseline. The baseline curves show the truck starts to heavily decelerate at 23 seconds when the driver sees the amber light 150 m away from the intersection, and stops fully at the intersection stop line at 35 seconds. However, with the help of the connected eco-driving system, the connected truck starts to accelerate prior to the signal turns to amber. Therefore, it can travel through the intersection without a full stop before the signal turns to red. Fig. \ref{Deceleration} shows the comparison results on the deceleration scenario between the connected eco-driving system and the baseline. The baseline curves show the truck starts to decelerate at 18 seconds while the driver sees the amber light 150 m away from the intersection, and stops fully at the intersection stop line at  28 second. However, with the help of the connected eco-driving system, the connected truck starts to decelerate prior to the signal turns to amber, and conducts a relatively smoother deceleration towards the intersection. Therefore, it can pass the intersection right after the signal turns to green with a low speed, instead of experiencing a full stop. We also adopt U.S. EPA's MOVES model to calculate the fuel consumption of aforementioned four test runs, where in each scenario the starting and ending speed are the same (Fig. \ref{Acceleration} and Fig. \ref{Deceleration} only show partial trajectories) \cite{MOVES}. The results of MOVES show that the connected eco-driving system can provide approximately 9\% and 4\% fuel savings in the acceleration and deceleration scenario, respectively.

Those four test runs shown above, along with many other test runs we have conducted, prove the proposed connected eco-driving system can smooth the speed variations of the truck, avoid unnecessary full stop at the intersection, and therefore reduce the fuel consumption of the heavy-duty truck. Another major finding of this field implementation is that, the cellular-based wireless communication approach enables CAVs to receive SPaT information from anywhere (not limited to a certain distance to the intersection), so the connected eco-driving can be implemented to a much larger scale. A disadvantage of this approach is an extra 100 ms communication delay compared to DSRC approximately, but this delay does not affect the behavior of this system to a large extent. The video of the field implementation could be watched online using the following link: https://www.youtube.com/watch?v=1CR4vMh8ufE.

\section{Conclusion and Future Work}
In this paper, a connected eco-driving system was developed for heavy-duty trucks traveling through cellular-based connected signalized intersections, aiming to smooth the speed profiles of the truck and hence reduce energy consumption. The system architecture was presented, introducing the different modules of the system. The map matching and eco-driving algorithms were described in detail. Field trials of the system with a heavy-duty diesel truck were conducted in Carson, California, showing the benefits of the system in real-world traffic.

Potential next steps include quantifying reductions in key criteria pollutant and GHG emissions, optimizing vehicle and system-level efficiencies, predicting queues at connected intersections, and automating the longitudinal speed and acceleration control of the connected truck.


\section*{Acknowledgment}
The vision of a connected vehicle test bed for “Eco-Drive” brought together two independent projects funded by California Air Resources Board, California Energy Commission, and South Coast Air Quality Management District. The California Collaborative Advanced Technology Drayage Truck Demonstration Project is part of California Climate Investments, a statewide initiative that puts billions of Cap-and-Trade dollars to work reducing greenhouse gas emissions, strengthening the economy, and improving public health and the environment — particularly in disadvantaged communities. 

We greatly acknowledge the equipment and technical support from Econolite, McCain, and Western Systems. We are also extremely grateful for the agency coordination and support received from Los Angeles County Metropolitan Transportation Authority, Los Angeles County Department of Public Works, City of Carson, and Port of Los Angeles. Finally, we extend thanks to Daniel Sandez, Anuja Patil, Dylan Brown, and Xishun Liao for their invaluable support during the field trials.



\bibliographystyle{IEEEtran}

%

\end{document}